\documentstyle[twocolumn,prl,floats,aps]{revtex}
\begin{document}
\twocolumn[\hsize\textwidth\columnwidth\hsize\csname @twocolumnfalse\endcsname
\title{Exact solutions and elementary excitations in the XXZ spin chain with unparallel boundary fields}
\author{Jun-peng Cao$^1$ \and Hai-Qing Lin$^2$\and  Kang-jie Shi$^3$ \and Yupeng Wang$^{1,4}$}
\address{$^1$Institute of Physics \& Center for Condensed Matter Physics, Chinese Academy of Sciences, 
Beijing 100080, People's Republic of China\\
$^2$Department of Physics, Chinese University of Hong Kong, Hong Kong, People's Republic of China\\
$^3$Institute of Modern Physics, Northwest University, Xi'an 710069, People's Republic of China\\
$^4$International Center for Quantum Structures, Chinese Academy of Sciences, Beijing 100080, People's Republic of China}  
\maketitle
\date{\today}
\maketitle
\begin{abstract}
By using a set of gauge transformations, the exact solutions of the XXZ spin chain with unparallel boundary magnetic 
fields are derived in the framework
of the algebraic Bethe ansatz. In the easy-plane case, we show
the elementary excitations 
are some kind of spinons without definite spin because the $U(1)$ symmetry is broken, while in the easy-axis ferromagnetic case a spiral state is 
realized in the ground state.
The correlation functions, the spin torque as well as the  spin voltage for the later case are also derived.
\end{abstract}
]
\narrowtext
The magnetic structure in a variety of condensed matters has been an important issue 
in modern physics. In lower dimensions, the situation is especially
distinctive and the spins may
show a variety of exotic features due to the strong quantum fluctuations. 
For instance, the elementary excitations in the one-dimensional (1D) 
spin liquids are doublet kinks\cite{fad} rather than triplet spin 
waves or magnons. Such quanta carry spin 1/2 and are now known as 
spinons. Furthermore, in spintronics, spins may carry 
information so that extended capability and performance can be added 
into the electronic device.\cite{wol} Several ways to produce pure 
spin current (without electric current) have been proposed, such as spiral spin structure as a 
driving force,\cite{mcd} spin Hall effect in ferromagnetic 
metals,\cite{hir} and ferromagnetic resonance.\cite{bra,ber} However, so 
far no experimental realization of pure spin current has been 
achieved. The spiral structure is a long range ordered state and 
therefore could hardly  exist in 1D materials such as quantum wires. Some 
interesting questions one may ask are: Can two unparallel boundary magnetic fields 
generate a spiral spin state in 1D? How does a spin liquid response 
to such fields? Since generally these fields break 
the $U(1)$ symmetry of the system, what is the nature of elementary 
excitations in 1D spin liquid under such fields? 
\par
As proposed in Ref.[3], the spin-supercurrent state
may occur in easy-plane ferromagnets. In this work, we show that
the spin structure, and consequently, the spin voltage may be 
controlled by the magnetic anisotropy represented by the
boundary magnetic fields.  By using a set of local transformations, the initial 
state for algebraic Bethe ansatz is established and the exact 
eigen-states of the model Hamiltonian are constructed. In the 
easy-plane case, the elementary excitations are almost the same as the 
usual spinons in spectrum but show spiral behavior in the real space 
and do not carry definite spin because the $U(1)$-symmetry is broken. In the easy-axis 
ferromagnetic case, the ground state is a spiral state. The magnetization, the correlation functions, the spin torque as well 
as the spin voltage in the ground state are derived exactly. The model 
Hamiltonian we shall consider reads 
\begin{eqnarray}
H&=&\sum_{j=1}^{N-1}(\sigma_j^x\sigma_{j+1}^x+\sigma_j^y\sigma_{j+1}^y+\cos\eta\sigma_j^z\sigma_{j+1}^z)\nonumber\\
&+&\vec{h}_1\cdot\vec{\sigma}_1
+\vec{h}_2\cdot\vec{\sigma}_N,
\end{eqnarray}
where $\sigma_j^\alpha$ ($\alpha=x,y,z$) are the Pauli matrices, $\eta$ is the anisotropic parameter and $\vec{h}_1=C_1(\sin\tilde{\theta}, \cos\tilde{\theta}, h_1)$
and $\vec{h}_2=C_2(\sin\theta, \cos\theta, h_2)$ are the boundary fields. If both $\vec{h}_{1,2}$ along the z-direction of the spin space, 
the system possesses $U(1)$
symmetry, i.e., the z-component of the total spin is conserved and the elementary excitations are the usual spinons.\cite{fad}. However, if one of the boundary
fields has non-zero transverse component, the $U(1)$ symmetry is broken and the spin structure may be different.
 We note that the present model has a lot of important applications in several physical systems. It has
been providing valuable insights into non-equilibrium behavior and complex dynamics\cite{gwa,sch,stin1,stin2} and broken symmetry.\cite{bar}
Interestingly, this model
has been proven to be integrable\cite{bat} via the open boundary transfer matrix method.\cite{skl} However, the spectrum  has not 
been derived via the conventional Bethe ansatz
scheme and leaves as an open problem in the field of integrable models. The difficulty comes from the absence of an obvious initial state or pseudo vacuum
(Without the boundary fields, it is the state where all spins polarized along the z-direction), thus the known Bethe ansatz can not be used directly.

In the following text, by using a set of local gauge transformations, we shall construct an adequate initial state with which the eigen-states 
can be derived via the algebraic Bethe ansatz. To show our procedure clearly, let us review the integrability of the model (1) briefly.
 The integrability of the XXZ spin chain is associated with the 
Lax matrix
\begin{eqnarray}
L_{0n}(u)&=&\frac 1{2\sin\eta}(\sin(u+\eta)+\sin u)\nonumber\\
&+&\frac1{2\sin\eta}(\sin(u+\eta)-\sin u)\sigma_0^z\sigma_n^z\\
&+&\sigma_0^+\sigma_n^-+\sigma_0^-\sigma_n^+,\nonumber
\end{eqnarray}
which satisfies the Yang-Baxter relation\cite{yan,bax}
\begin{eqnarray}
R_{\alpha\beta}L_{\alpha n}(u)L_{\beta n}(v)=L_{\beta n}(v)L_{\alpha n}(u)R_{\alpha\beta},
\end{eqnarray}
with $R_{\alpha\beta}\equiv L_{\alpha\beta}(u-v)$.
The monodromy matrix 
$T_0(u)=L_{01}(u)\cdots L_{0N}(u)$ also satisfies the Yang-Baxter relation.
In addition, as demonstrated by Sklyanin\cite{skl}, the so called reflection equation 
\begin{eqnarray}
R_{\alpha\beta}(u-v)K_\alpha^-(u)R_{\alpha\beta}(u+v)K_\beta^-(v)\nonumber\\
=K_\beta^-(v)R_{\alpha\beta}(u+v)K_\alpha^-(u)R_{\alpha\beta}(u-v)
\end{eqnarray} 
is also necessary for establishing the open boundary integrable models. Since the double-row monodromy matrix
\begin{eqnarray}
U_0(u)=T_0(u)K_0^-(u)T_0^{-1}(-u)
\end{eqnarray}
also satisfies the reflection equation, the transfer matrices $t(u)=tr_0K_0^+(u)U_0(u)$ with different spectral parameters are mutually commutative, i.e.,
 $[t(u), t(v)]=0$ as long as the dual reflection 
matrix $K_0^+(u)$ satisfies the dual reflection equation.\cite{skl} Therefore, $t(u)$ serves as the generation function of the conserved 
quantities. For the XXZ spin chain, the general solution of $K_0^\pm(u)$ read\cite{dev,bat}
\begin{eqnarray}
K_0^-(u)&=&c_{00}\cos u-c_{01}\sin u\sigma_0^z\nonumber\\
&+&\sin(2u)(\sin\theta \sigma_0^x+\cos\theta\sigma_0^y),\nonumber\\
K_0^+(u)&=&-\tilde{c}_{00}\cos(u+\eta)-\tilde{c}_{01}\sin(u+\eta)\sigma_0^z\\
&+&\sin2(u+\eta)(\sin\tilde{\theta}\sigma_0^x+\cos\tilde{\theta}\sigma_0^y),\nonumber
\end{eqnarray}
where $c_{00}$, $c_{01}$, $\tilde{c}_{00}$, $\tilde{c}_{01}$, $\theta$ and $\tilde{\theta}$ are arbitrary parameters. The Hamiltonian can be
expressed as
\begin{eqnarray} 
H=\sin\eta\frac{d}{du}\ln t(u)|_{u=0}-N\cos\eta,
\end{eqnarray}
with $C_1=-2\sin\eta\tilde{c}^{-1}_{00}$, $h_1=-\tilde{c}_{01}/2$, $C_2=2\sin\eta c_{00}^{-1}$ and $h_2=-c_{01}/2$.  
\par
An obvious initial state exists in the conventional algebraic Bethe ansatz scheme, acting on which, the lower left element of the 2$\times$2 matrix $U_0(u)$
 is zero. However, in the present case, the initial state is not obvious because $K_0^{+}$ and $K_0^{-}$ are generally not diagonal. To find a proper initial state,
as used in solving the eight-vertex model,\cite{bax,tak,fan} we introduce two sets of gauge matrices $M_n(u)$ and $\bar{M}_n(u)$. Under the gauge transformations
\begin{eqnarray}
\bar{L}_{0n}(u|m)&=&\bar{M}^{-1}_{n-1+m}(u)L_{0n}(u)\bar{M}_{n+m}(u),\nonumber\\
S_{0n}(-u|m)&=&M_{n+m}^{-1}(-u)L_{0n}^{-1}(-u)M_{n-1+m}(-u),
\end{eqnarray}
the lower left elements of $\bar{L}_{0n}$ and its inverse $S_{0n}$ acting on a proper local state is zero. 
 The above requirements define the forms of the gauge matrices as
\begin{eqnarray}
M_n(u)=\left( X_n(u), Y_n(u)\right),\nonumber\\
\bar{M}_n(u)=\left( X_{n+1}(u), Y_{n-1}(u)\right),
\end{eqnarray}
where $X_n$ and $Y_n$ are 1$\times$2 column matrices with upper elements $e^{-iu}x_n$ and $e^{-iu}y_n$, respectively; their lower elements are defined
as unity. The constants $x_n$ and $y_n$ read
\begin{eqnarray}
x_n=-ie^{-i\eta(\beta+\gamma+n)},\;\; y_n=-ie^{-i\eta(\beta-\gamma-n)},
\end{eqnarray}
where $\beta$ and $\gamma$ are two parameters determined by the boundary fields. In addition, we define
\begin{eqnarray}
&&\bar{X}_n(u)=\frac 1{x_n-y_n}(-e^{iu}, x_n),\nonumber\\
&&\bar{Y}_n(u)=\frac1{x_n-y_n}(e^{iu},-y_n),\nonumber\\
&&\tilde{X}_n(u)=\frac 1{x_n-y_{n-2}}(-e^{iu}, x_n),\nonumber\\
&&\tilde{Y}_n(u)=\frac1{x_{n+2}-y_n}(e^{iu},-y_n). 
\end{eqnarray}
In this case, the local vacuum of site $n$ reads
\begin{eqnarray}
|\omega>_n^m=x_{n+m}|\uparrow>_n+|\downarrow>_n.
\end{eqnarray}
It is easily to show that the lower left elements of $S_{0n}(-u|m)$ and ${\bar L}_{0n}(u|m)$ acting on this state is zero, while the upper
diagonal element of $S_{0n}(-u|m)$ and the lower diagonal element of ${\bar L}_{0n}(u|m)$ shift $|\omega>_n^m$ to $|\omega>_n^{m-1}$, and
the lower diagonal element of $S_{0n}(-u|m)$ and the upper diagonal element of ${\bar L}_{0n}(u|m)$ shift $|\omega>_n^m$ 
to $|\omega>_n^{m+1}$, respectively.   
Now we can rewrite the transfer matrix as
\begin{eqnarray}
t(u)=tr_0\tilde{K}_0^+(u|m_0)\tilde{U}_0(u|m_0),
\end{eqnarray}
where 
\begin{eqnarray}
&&\tilde{K}_0^+(u|m)=\left(
\begin{array}{cc}
\tilde{K}_0^+(u|m)_{11} & \tilde{K}_0^+(u|m)_{12} \\
\tilde{K}_0^+(u|m)_{21} & \tilde{K}_0^+(u|m)_{22}
\end{array}
\right)=\nonumber\\
&&\left(
\begin{array}{cc}
{\bar Y}_{m}(-u)K^{+}(u)X_{m}(u)  &
{\bar Y}_{m}(-u)K^{+}(u)Y_{m-2}(u) \\
{\bar X}_{m}(-u)K^{+}(u)X_{m+2}(u)  &
{\bar X}_{m}(-u)K^{+}(u)Y_{m}(u)
\end{array}
\right),\nonumber
\end{eqnarray}
and 
\begin{eqnarray}
&&\tilde{U}_0(u|m)=\left(
\begin{array}{cc}
A_m(u) & B_m(u) \\
C_m(u) & D_m(u)
\end{array}
\right)=\nonumber\\
&&\left(
\begin{array}{cc}
 {\tilde Y}_{m-2}(u)U(u)X_{m}(-u)  &
{\tilde Y}_{m}(u)U(u)Y_{m}(-u) \\
{\tilde X}_{m}(u)U(u)X_{m}(-u)  &
{\tilde X}_{m+2}(u)U(u)Y_{m}(-u)
\end{array}
\right).
\end{eqnarray}
To ensure  $\tilde{K}_0^+(u|m_0)$ being diagonal, we
parameterize the left boundary field as
${\tilde c}_{00}+i{\tilde c}_{01}=-2\cos(m_0+\gamma)\eta$,
then
\begin{eqnarray}
t(u)=\tilde{K}_0^+(u|m_0)_{11}A_{m_0}(u)+\tilde{K}_0^+(u|m_0)_{22}D_{m_0}(u).
\end{eqnarray}
In addition, $C_m(u)$ must annihilate the global initial state
$|\Omega>^m=\Pi_\otimes|\omega>_n^m$. This needs $c_{00}+ic_{01}=-2\cos(\theta-\tilde \theta+\eta(\gamma+N+m-1))$,
which gives a constraint of $c_{00}, c_{01}, {\tilde c}_{00}$ and ${\tilde c}_{01}$. Under the above condition, $|\Omega>^m$
is a common eigen-state of $A_m(u)$ and $D_m(u)$, and
$B_m(u)$ serve as the generating operators of the eigen-states of $t(u)$ as usual. To derive the Bethe states, we need also
the commutation relations between  $A_m$, $D_m$ and $B_n$. With the help of the production relations of 
$R_{1 2}(u_1-u_2)X^{1}_{m+2}(u_1)X^{2}_{m+1}(u_2)$ etc.\cite{tak} and the reflection equation  of $U_0(u)$, 
after some manipulation we obtain
\begin{eqnarray}
B_m(u_1)B_{m-2}(u_2)=B_m(u_2)B_{m-2}(u_1),\\
A_{m+2}(u_1)B_{m}(u_2)=\frac{\sin(u_1+u_2)\sin(u_1-u_2-\eta)}
{\sin(u_1-u_2)\sin(u_1+u_2+\eta)}\nonumber\\
\times
 B_{m}(u_2)A_{m}(u_1)
\nonumber\\-\frac{\sin(2u_2) \sin\eta \sin(u_1-u_2-\eta(m+\gamma+1))}
             {\sin(u_1-u_2) \sin\eta(m+\gamma+1) \sin(2u_2+\eta)}
\nonumber\\
\times B_{m}(u_1)A_{m}(u_2)
\nonumber\\-\frac{\sin(-u_1-u_2+\eta(m+\gamma))\sin\eta \sin\eta}
             {\sin\eta(m+\gamma+1) \sin(u_1+u_2+\eta)\sin(2u_2+\eta)}\nonumber\\
             \times
B_{m}(u_1){\tilde D}_{m}(u_2),\\
{\tilde D}_{m+2}(u_1)B_{m}(u_2)=\nonumber\\
\frac{\sin(u_1-u_2+\eta)\sin(u_1+u_2+2\eta)}
{\sin(u_1-u_2)\sin(u_1+u_2+\eta)} \nonumber\\
\times B_{m}(u_2){\tilde D}_{m}(u_1) 
\nonumber \\
-\frac{\sin(u_1-u_2+\eta(m+\gamma+1))\sin(2u_1+2\eta) \sin\eta}
             {\sin\eta(m+\gamma+1) \sin(u_1-u_2)\sin(2u_2+\eta)}\nonumber\\
             \times
B_{m}(u_1){\tilde D}_{m}(u_2)  
\nonumber\\
+\frac{\sin(2u_2)
\sin(u_1+u_2+\eta(m+\gamma+2))}
{\sin(u_1+u_2+\eta) \sin\eta(m+\gamma+1) }\nonumber\\
\times\frac{\sin(2u_1+2\eta)}{\sin(2u_2+\eta)} B_{m}(u_1)A_{m}(u_2),
\end{eqnarray}
where
\begin{eqnarray}
{\tilde D}_{m}(u)=\frac{\sin(2u+\eta)\sin\eta(m+\gamma+1)}
{\sin\eta \sin\eta(m+\gamma)}
D_{m}(u)\nonumber\\
-\frac{\sin(\eta(m+\gamma)+2u+\eta)}{\sin\eta(m+\gamma)}
A_{m}(u).
\end{eqnarray}
\par
Assume the eigen-state of the transfer matrix take the form
\begin{eqnarray}
|\Phi>=B_{m_0-2}(u_1)\cdots B_{m_0-2M}(u_M)|\Omega>^m ~,
\end{eqnarray}
with $m_0-2M=m$.
Acting the transfer matrix on this state and exchanging the position of $A_{m_0}(u)$, $\tilde{D}_{m_0}(u)$ and
$B_{m_0-2}(u_1)B_{m_0-4}(u_2)\cdots B_{m_0-2M}(u_M)$ by the commutation relations, we obtain two types of terms.
One gives the eigen value of $t(u)$ while in the other type one spectral parameter $u_j$
exchanges with $u$.
The latter terms are set to zero and we obtain the Bethe ansatz equations
\begin{eqnarray}
\frac{\sinh^{2N}(\lambda_j-\frac{1}{2}i\eta)}{\sinh^{2N}(\lambda_j+\frac{1}{2}i\eta)}=\nonumber\\
\frac{
\sinh(\lambda_j+{ i{\tilde \alpha}}_1-\frac{1}{2}i\eta)
\cosh(\lambda_j+{ i{\tilde \alpha}}_2-\frac{1}{2}i\eta)}{
\sinh(\lambda_j-{ i{\tilde \alpha}}_1+\frac{1}{2}i\eta)
\cosh(\lambda_j-{i{\tilde \alpha}}_2+\frac{1}{2}i\eta)}\nonumber\\
\times\frac{\sinh(\lambda_j-{i\alpha}_1-\frac{1}{2}i\eta)
\cosh(\lambda_j-{i\alpha}_2-\frac{1}{2}i\eta)}{\sinh(\lambda_j+{i\alpha}_1+\frac{1}{2}i\eta)
\cosh(\lambda_j+{i \alpha}_2+\frac{1}{2}i\eta)}\nonumber\\
\times
\prod_{l\neq j}^{M}
\frac{\sinh(\lambda_{j}+\lambda_l-i\eta)\sinh(\lambda_{j}-\lambda_l-i\eta)}
{\sinh(\lambda_j+\lambda_{l}+i\eta)\sinh(\lambda_j-\lambda_{l}+i\eta)},
\nonumber\\
j=1,2,\cdots,M.
\end{eqnarray}
The energy spectrum is given by
\begin{eqnarray}
E=(N-1)\cos\eta 
-\frac{
\sin\eta \sin({ {\tilde \alpha}}_1-{ \alpha}_1)}
{\sin { {\tilde \alpha}}_1 \sin  {\alpha}_1}\nonumber\\
-\sum_{j=1}^{M}
\frac{2\sin^2 \eta}{\cosh^2\lambda_j-\cos^2(\frac 12\eta)},
\end{eqnarray}
with the parameterizations
$
c_{00}=-2\sin\alpha_1 \cos\alpha_2,~~~ 
c_{01}=2i\cos\alpha_1 \sin\alpha_2,~~~
\tilde c_{00}=-2\sin \tilde \alpha_1 \cos \tilde \alpha_2,~~~
\tilde c_{01}=2i\cos \tilde \alpha_1 \sin \tilde \alpha_2.
$
Here we have put $u_j\to i\lambda_j-\eta/2$.
The Bethe ansatz equations look very similar to that of the usual case. However, there is a restriction on the number of ``flipped'' spins $M$.
In fact,  the two boundary fields are not independent with each other but related with a complex parameter $\gamma+m_0$ and an integer $M$.
That means given a right boundary
field, the left boundary is restricted by the right one and $M$ or vise versa. In addition, by exchanging the two fields, we can obtain another set of
solutions with $M\to N-M-1$ and the same Bethe ansatz equations and spectrum.  We choose
 ${\tilde c}_{00}+i{\tilde c}_{01}=-2\cos(m_0+\gamma)\eta$
 and  $c_{00}+ic_{01}=-2\cos(\theta-\tilde \theta+\eta(m_0+\gamma-1))$ for real $\eta$ (easy-plane case) and even $N$.
In this case, $M=N/2,\;N/2-1$. This allows us to study the ground state and the low lying excitation properties simultaneously. 
In the charge neutral sector ($M=N/2$), the ground state is described by real $\lambda_j$ which fill the whole real axis 
in the continuum limit;
 the excitations are $\lambda-n$-strings with $n$-holes in the real axis. In the thermodynamic limit $N\to\infty$, the excitation energy reads\cite{bog}
 \begin{eqnarray}
 \epsilon(\lambda_1^h,\cdots\lambda_n^h)=\sum_{j=1}^n\epsilon(\lambda_j^h)=\sum_j^n\frac{\pi\sin \eta}{\eta\cosh\frac{\pi\lambda_j^h}{2\eta}},
 \end{eqnarray}
 i.e., the spectrum takes the same form
of the case without the boundary fields, where $\lambda_j^h$ is the position of $j$-th hole in the spectral space and $\epsilon(\lambda_j^h)$
is the energy of a single spinon.\cite{bog} The other type of excitations can be constructed by considering the process of $M=N/2$ to $M=N/2-1$, i.e.,
 a two-hole state based on the ground state. The excitation energy simply reads
 \begin{eqnarray}
 \epsilon(\lambda_1^h,\lambda_2^h)=\epsilon(\lambda_1^h)+\epsilon(\lambda_2^h),
 \end{eqnarray}
 where $\lambda_{1,2}^h$ are the positions of the two holes in the spectral space. This state corresponds to the spin triplet excitation
  in the usual case. The above discussion shows that the elementary excitations are not changed by the boundary fields in spectrum, i.e, they
  are also described by spinons formally (additive spectrum).
Since the spinons are some kind of topological excitations, it is rather
strange that the broken symmetry does not alter the spectrum.
However, the spin structure in the real space is indeed changed by
the boundary fields.
The initial state is in fact a spiral state with the spins rotate site by site in the $x-y$ plane. The ``spin-wave'' generators $B_m(u)$ do not 
``flip'' the spins on different sites to an uniform direction. Therefore the spins have winding behavior around the $z$-direction.

For the easy-axis ferromagnetic case, we put $\eta=\pi+i\chi$ with $\chi$ real and study the
$\tilde{c}_{00}+i\tilde{c}_{01}=-2\cos((\pi+i\chi)(m_0+\gamma))$,
and $c_{00}+ic_{01}=-2\cos(\theta-\tilde \theta +(\pi+i\chi)(m_0+\gamma+N-1))$ case.
 The ground state is described by the reference state $|\Omega>^{m_0}$.
 Several quantities in the ground state can be derived easily. For instance,
the z-direction magnetization is
\begin{eqnarray}
<M_z>= \frac12+O(N^{-1}),
\end{eqnarray}
implying that the boundary fields  change the magnetization only in order of $1/N$. The correlation functions can also be derived as
\begin{eqnarray}
<\sigma_n^z \sigma_m^z>=\tanh f(n)\tanh f(m),\nonumber\\
<\sigma_n^x \sigma_m^x>=\frac{(-1)^{n+m}\cos^2(g)}{\cosh f(n)\cosh f(m)},\\
<\sigma_n^y \sigma_m^y>=\frac{(-1)^{n+m}\sin^2(g)}{\cosh f(n)\cosh f(m)},\nonumber
\end{eqnarray}
with $f(n)=(Re(m_0+\gamma)+n-1)\chi+\pi Im(m_0+\gamma)$ and $g=\pi Re(m_0+\gamma)-\chi Im(m_0+\gamma)-\tilde \theta$.
Obviously, the z-z correlation function tends to 1 when both $m$ and $n$ tend to infinity, indicating the long-range order of the ground state.
Define the quantum spin torque operator as
${\vec S}_n=\frac 14{\vec\sigma}_n\times{\vec\sigma}_{n+1}$.
The expectation value of this quantity in the ground state reads
\begin{eqnarray*}
<S_n^x>&=&(-1)^{n}\frac{\sin(g)\sinh[f(n)+\chi/2]\cosh(\chi/2)}{2\cosh f(n+1)\cosh f(n)},\nonumber\\
<S_n^y>&=&(-1)^{n+1}\frac{\cos(g)\sinh[f(n)+\chi/2]\cosh(\chi/2)}{2\cosh f(n+1)\cosh f(n)},\\
<S_n^z>&=&0. \nonumber
\end{eqnarray*}
The spin voltage can also be calculated   as $V_s^\alpha=\sum_{n=1}^{N-1}<S_n^\alpha>$, $\alpha=x,y,z$.

In conclusion, we develop a method to derive the eigen-states of the XXZ spin chain with unparallel boundary fields. For some special
choices of the boundary fields, the ground state and the low lying excitations
for the easy-plane case are obtained exactly.
The spectrum of the excitations is of the same form as that of the usual spinons without definite charge (spin),
suggesting that there exist some hidden symmetries in the system,
an interesting question needs to be further explored.
 For the easy-axis ferromagnetic case, the ground state is a pure spiral state  which possesses
non-zero  spin torque in the x-y plane. This provides a possible scheme to generate driving force for spin current in quasi 1D systems.

One of the authors (Y. Wang) was supported by the NSF of China. He also thanks Physics Department of CUHK for their hospitality during
his visit and acknowledges financial support from project CUHK 4246/01P.


\begin{references}
\bibitem{fad} L.D. Faddeev and L.A. Takhtajian, Phys. Lett. {\bf 83A}, 375 (1981).
\bibitem{wol} A. A. Wolf {\it et al.}, Science {\bf 294}, 1488 (2001).
\bibitem{mcd} J. Konig, M. C. Bosager, and A. H. MacDonald, Phys. Rev.
Lett. {\bf 87}, 187202 (2001).
\bibitem{hir} J. E. Hirsch, Phys. Rev. Lett. {\bf 83}, 1834 (1999); Phys. Rev. B
{\bf 60}, 14787 (1999).
\bibitem{bra} A. Brataas, Y. Tserkovnyak, G. E. W. Bauer, and B. I.
Halperin, cond-mat/0205028 (unpublished).
\bibitem{ber} L. Berger, Phys. Rev. B {\bf 59}, 11465 (1999); J. Appl. Phys.
{\bf 90}, 4632 (2001).
\bibitem{gwa} L.-H. Gwa and H. Spohn, Phys. Rev. Lett. {\bf 68}, 725 (1992).
\bibitem{sch} G. Sch\"{u}tz, J. Stat. Phys. {\bf 71}, 471 (1993).
\bibitem{stin1} R.B. Stinchcombe and G.M. Sch\"{u}tz, Europhys. Lett. {\bf 29}, 663 (1995).
\bibitem{stin2} R.B. Stinchcombe and G.M. Sch\"{u}tz, Phys. Rev. Lett. {\bf 75}, 140 (1995).
\bibitem{bar} M. Barma, M.D. Grynberg and R.B. Stinchcombe, Phys. Rev. Lett. {\bf 70}, 1033 (1993);
 R.B. Sinchcombe, M.D. Grynberg and M. Barma, Phys. Rev. E {\bf 47}, 4018 (1993).
\bibitem{bat} M.T. Batchelor, {\it Proceedings of the 22nd International Colloquium on Group Theoretical Methods in Physics}, 
eds S.P. Corney et al, 261 (International Press, Boston, 1999),
cond-mat/9811165.
\bibitem{skl} E.K. Sklyanin, J. Phys. A {\bf 21}, 2375 (1988).
\bibitem{yan} C.N. Yang, Phys. Rev. Lett. {\bf 19}, 1312 (1967).
\bibitem{bax} R.J. Baxter, Phys. Rev. Lett. {\bf 26}, 832 (1971); {\bf 26}, 834 (1971);
Ann. Phys. {\bf 80}, 193; 323 (1972).
\bibitem{dev} H.J. de Vega and A. Gonzalez-Ruiz, J. Phys. A {\bf 26}, L519 (1993);
{\bf 27}, 6129 (1994).
\bibitem{tak} L.A. Takhtajian and L.D. Faddeev, Russian Math. Surveys {\bf 34:5}, 13 (1979). 
\bibitem{fan} H. Fan, B.-Y. Hou, K.-J. Shi and Z.-X. Yang,
Nucl. Phys. B {\bf 478}, 723 (1996).
\bibitem{bog} For example, see, V.E. Korepin, N.M. Bogoliubov and A.G. Izergin, {\it Quantum Inverse Scattering Method and 
Correlation Functions}, (Cambridge University Press, Cambridge, 1993).


\end{references}
\end{document}